\documentclass[epj]{svjour}
\usepackage{latexsym,graphicx,epsfig,psfrag,here}
\newcommand{\beq}{\begin{equation}} 
\newcommand{\eeq}{\end{equation}} 
\newcommand{\beqa}{\begin{eqnarray}} 
\newcommand{\eeqa}{\end{eqnarray}} 
\newcommand{\beqan}{\begin{eqnarray*}} 
\newcommand{\eeqan}{\end{eqnarray*}} 
\newcommand{\ba}{\begin{array}} 
\newcommand{\ea}{\end{array}} 
\newcommand{\no}{\nonumber}

\newcommand{\ve}{\varepsilon}

\newcommand{\wt}{\widetilde} 
\newcommand{\wh}{\widehat}

\newcommand{\D}{{\cal D}}

\newcommand{\cL}{{\cal L}}

\newcommand{\cO}{{\cal O}} 
 
\newcommand{\Q}{{\cal Q}}

\newcommand{\nn}{\nonumber \\}

\newcommand{\bea}{\begin{eqnarray}} 
\newcommand{\eea}{\end{eqnarray}}

\begin{document}
\onecolumn
\thispagestyle{empty}
\begin{flushright} 
IFIC/03-41\\
MAP-293\\
UWThPh-2003-22\\ 
PSI-PR-03-13\\ 
\end{flushright} 
\vspace{2.0cm} 
\begin{center} 
{\Large \bf $\mbox{\boldmath $K_{e3}$}$ decays and CKM unitarity~*}\\[40pt] 
{\large V. Cirigliano$^{1,2}$, H. Neufeld$^3$, H. Pichl$^4$} 
 
\vspace{1cm} 
$^1$ Departament de F\'{\i}sica Te\`{o}rica, IFIC, Universitat 
de Val\`{e}ncia -- CSIC, \\
Apt. Correus 22085, E-46071 Val\`{e}ncia, Spain \\[10pt]

$^2$ Department of Physics, California Institute of Technology, Pasadena, 
California 91125, USA\\[10pt]

$^3$ Institut f\"ur Theoretische Physik, Universit\"at 
Wien, Boltzmanngasse 5, A-1090 Wien, Austria \\[10pt]

$^4$ Paul Scherrer Institut, CH-5232 Villigen PSI, Switzerland\\

\vspace*{2.0cm}
{\bf Abstract} \\ 
\end{center} 
\noindent We present a detailed numerical study of the $K_{e3}$ decays to $\cO(p^6, 
(m_d-m_u)p^2, e^2 p^2)$ in chiral perturbation theory with virtual photons and leptons. 
We describe the extraction of the CKM matrix element $|V_{us}|$ from the experimental 
$K_{e3}$ decay parameters. We propose a consistency check of the $K^+_{e3}$ and $K^0_{e3}$ 
data that is largely insensitive to the dominating theoretical uncertainties, in particular 
the contributions of $\cO(p^6)$. Our analysis is highly relevant in view of the recent high 
statistics measurement of the $K^+_{e3}$ branching ratio by E865 at Brookhaven which does 
not indicate any significant deviation from CKM unitarity but rather a discrepancy with the 
present $K^0_{e3}$ data.   
\vfill 
\noindent *~Work supported in part by IHP-RTN, Contract No. HPRN-CT2002-00311 
(EURIDICE) and by Acciones Integradas, Project No. 19/2003 (Austria), HU2002-0044 (MCYT, 
Spain)
\newpage 
\twocolumn
%
%
%
\section{Introduction}
\label{sec: Introduction}
\renewcommand{\theequation}{\arabic{section}.\arabic{equation}}
\setcounter{equation}{0}

According to the compilation of the Particle Data Group (PDG) 2002 
\cite{PDG02}, the absolute values of the entries in the first row of the 
Cabibbo--Kobayashi--Maskawa (CKM) mixing matrix are given by 
\beqa \label{CKM}
 |V_{ud}| &=& 0.9734 \pm 0.0008  , \no \\
 |V_{us}| &=& 0.2196 \pm 0.0026  , \no \\
 |V_{ub}| &=& 0.0036 \pm 0.0010  , 
\eeqa
which implies a $2.2 \, \sigma$ deviation from  unitarity:  
\beq \label{unitarity1}
|V_{ud}|^2 + |V_{us}|^2 + |V_{ub}|^2 - 1 = -0.0042 \pm 0.0019.
\eeq    
The value for $|V_{ud}|$ in (\ref{CKM}) has been extracted from 
super-allowed Fermi transitions of several $0^+$ nuclei and neutron beta 
decay, whereas  the number for $|V_{us}|$ is based on more than 
thirty-year-old $K_{e3}$ data.  

The situation has changed dramatically with the outcome of 
a new high statistics measurement of the  
$K^+_{e3}$ branching ratio by the E865 
Collaboration at Brookhaven \cite{E865}. Their analysis of more than 
70,000 $K^+_{e3}$ events yielded a  branching ratio  
which was about $2.3 \, \sigma$ larger than the current PDG value. 
As a consequence, the value of $|V_{us}|$ based on the new
experimental result does not indicate any significant deviation from
unitarity.  
Moreover, besides indicating a sharp disagreement between new and old
$K^+_{e3}$ data, the new result implies 
an inconsistency between $K^+_{e3}$ and $K^0_{e3}$ data.  
 
The current experimental information on the decay mode of the neutral 
kaon is 
indeed very unsatisfactory. The two numbers given by PDG 2002 \cite{PDG02},
\beqa \label{K0e3data}
\Gamma(K^0_{e3})_{\rm fit} &=& (7.50 \pm 0.08) 
\times 10^6 \, {\rm s^{-1}}, \no \\
\Gamma(K^0_{e3})_{\rm average} &=& (7.7 \pm 0.5) 
\times 10^6 \, {\rm s^{-1}}, 
\eeqa
differ considerably depending on the procedure for the 
treatment of data. The first value in (\ref{K0e3data}) was obtained from a 
constrained fit using all significant measured $K_L$ branching ratios, the 
second one is a weighted average of measurements of the $K^0_{e3}$ ratio 
only. Apparently, the rate obtained from the fit is completely driven by 
input 
different from the actual measurements. In particular the error on the 
``fitted'' value does not reflect at all the experimental accuracy (the 
experiments were made in the sixties and early seventies) but rather the 
constraints from the global fit.

Presently, new independent $K_{e3}$ decay measurements are in progress 
(CMD2, NA48, KLOE) and should help to clarify the experimental situation.

In this paper, we present a detailed numerical analysis 
of the radiative corrections to the $K^0_{e 3}$ Dalitz plot 
distribution. We discuss possible strategies to extract $|V_{us}|$ 
from the experimental data and we propose a rather powerful consistency 
check of $K^+_{e3}$ and $K^0_{e3}$ measurements.

This work is based on our previous calculation \cite{CKNRT02}
of the $K_{\ell 3}$ decays to $\cO(p^4,(m_d-m_u)p^2,e^2p^2)$ in chiral 
perturbation theory with virtual photons and leptons \cite{lept}.
After a brief review of the main kinematic features of $K_{e3}$ decays
and the structure of radiative corrections (Sect. \ref{sec: Kinematics}), 
we recall the structure of the form factors relevant for $K_{e3}$ decays 
including a discussion of the recent results on the contributions of 
order $p^6$ in the chiral expansion in Sect. \ref{sec: form factors}.
Real photon emission in the $K^0_{e 3}$ case is discussed in Sect. 
\ref{sec: Real photon radiation}. 
In Sect. \ref{sec: Extraction} we illustrate our general considerations
by a numerical study of the $K_{e 3}^0$ decay and the description of 
a procedure to extract the 
CKM matrix element $|V_{us}|$ from experimental 
data.
The impact of the E865 experiment on the determination of 
$|V_{us}|$ from $K^+_{e3}$ data is discussed in Sect. \ref{sec: E865}. 
A specific strategy for a combined analysis of $K^0_{e3}$ and $K^+_{e3}$ 
data is proposed in Sect. \ref{sec: comb}.
Our conclusions are summarized in Sect. \ref{sec: Conclusions}, and three 
Appendices contain some technical material related to the calculation of 
loop contributions and real photon radiation.

\section{Kinematics and radiative corrections}
\label{sec: Kinematics} 
\renewcommand{\theequation}{\arabic{section}.\arabic{equation}}
\setcounter{equation}{0}

The generic $K_{e3}$ decay
\beq
K (p_K) \to \pi (p_\pi) \, e^+ (p_e) \, \nu_e (p_\nu) 
\eeq
can be described by a single form factor (usually denoted by $f_+$). 
A second form factor\footnote{See \cite{CKNRT02} for the general $K_{\ell 
3}$ form factor decomposition}, being also present in principle, enters 
only together with the   
tiny quantity $m_e^2/M_K^2 \simeq 10^{-6}$ in the formula for the Dalitz 
plot density. Therefore, these 
contributions are utterly negligible and 
the invariant amplitude (in the absence of radiative corrections)
can be simplified to
\beq
{\cal M} = 
\frac{G_{\rm F}}{\sqrt{2}} \, V_{us}^{*} \, l^{\mu} \, C \,  f_{+}^{(0)} 
(t) \, (p_K + p_\pi)_{\mu} ,
\label{basic1}
\eeq
where 
\beq
l^\mu = \bar{u} (p_\nu) \, \gamma^\mu \, (1 - \gamma_5) \, v (p_e) 
\eeq
denotes the weak leptonic current, and
\beq
\quad C = \left\{ \begin{array}{ll} 1 & \, \mbox{for} \ K^{0}_{e 3} \\
1/\sqrt{2} & \, \mbox{for} \ K^{+}_{e 3} \end{array} \right.  .  
\eeq
The form factor depends on the single kinematical variable $t = (p_K - 
p_\pi)^2$
and the superscript $(0)$ indicates the limit $e=0$.

The spin-averaged decay distribution $\rho(y,z)$ for $K_{e 3}$ 
depends on the two variables
\beq z = \frac{ 2 p_K \! \cdot \! p_\pi }{M_K^2} = 
\frac{2 E_{\pi}}{M_K}  ,  \quad y = \frac{ 2 p_K \! \cdot \! p_e 
}{M_K^2} =
\frac{2 E_{e}}{M_K}  , \eeq
where $E_\pi$ ($E_e$) is the pion (positron) energy in the 
kaon rest frame, and $M_K$ indicates the mass of the decaying kaon. 
Alternatively one may also use two of the Lorentz 
invariants 
\beq
t = (p_K - p_\pi)^2   , \quad u = (p_K - p_e)^2   , 
\quad s = (p_\pi + p_e)^2 
 . 
\eeq
Then the distribution  reads 
\beq
\rho^{(0)} (y,z) =  {\cal N} \,
 A_1^{(0)}(y,z) \,  |f_{+}^{(0)} (t)|^2 , 
\label{basic2}
\eeq
with  
\beq \label{expl}
{\cal N}  = C^2 \frac{G_{\rm F}^2 \, |V_{us}|^2 M_K^5}{128 \pi^3}  , 
\quad \Gamma = \int\limits_{\cal D}  dy \, dz \ \rho^{(0)} (y,z)  .  
\eeq
The kinematical density is given by 
\beqa
A_1^{(0)} (y,z) &=& 4 (z + y - 1) (1 - y) 
+ r_e (4 y + 3 z - 3) \no \\
&&{}- 4 r_\pi + r_e (r_\pi - r_e)    ,  
\label{kin0}
\eeqa
where
\beq
r_e = \frac{m_e^2}{M_K^2}  , \qquad r_\pi = 
\frac{M_{\pi}^2}{M_K^2}  .
\eeq
The boundaries of the domain of integration $\cal D$ (Dalitz plot) in 
(\ref{expl}) can 
be found in Sect. \ref{sec: Real photon radiation}. 

Virtual photon exchange as well as the contributions of the appropriate 
electromagnetic counterterms change the form factor \cite{CKNRT02},
\beq 
f_{+}^{(0)} (t) \to F_+ (t,v)  , 
\eeq
and the distribution (\ref{basic2}) has to be replaced with 
\beq
\varrho (y,z)=  {\cal N} \,
 A_1^{(0)}(y,z) \,  |F_{+} (t,v)|^2 . 
\label{newrho}
\eeq
The full form factor $F_+ (t,v)$ 
depends now also on a second kinematical  
variable $v$ as it cannot be interpreted anymore as the matrix element of 
a quark current between hadronic states. 
The variable $v$ is taken as $u = (p_K - p_e)^2$ for
$K_{e 3}^+$ and $s = (p_\pi + p_e)^2$ for $K_{e 3}^0$.
Diagrammatically, the dependence on the second variable is generated by 
one-loop graphs where a photon line connects the charged meson and the  
positron. 

The form factor $F_+(t,v)$ contains infrared singularities due to 
low-momentum virtual photons. They can be regularized by introducing a 
small photon mass $M_{\gamma}$. The dependence on an infrared cutoff 
reflects the fact that $F_+(t,v)$ cannot be interpreted as an observable 
quantity but has to be combined with the contributions from real photon 
emission to arrive at an infrared-finite result.

It is convenient to decompose $F_+(t,v)$ into a structure-dependent 
effective form factor $f_+(t)$ and a remaining part containing in 
particular the universal long-distance corrections \cite{CKNRT02}. To 
order $\alpha$, the full form factor is given by 
\beq 
F_+ (t,v) = \left[ 1 + \frac{\alpha}{4 \pi} 
\Gamma (v,m_e^2,M^2;M_\gamma) \right] \ f_+ (t) , 
\label{factor1}
\eeq 
where $M$ denotes the mass of the charged meson.
Expressed in terms of the functions $\Gamma_c$, $\Gamma_1$, $\Gamma_2$ 
defined in \cite{CKNRT02}, $\Gamma$ can be written as 
\beqa \label{Gamma}
\Gamma(v,m_e^2,M^2;M_\gamma) &=& 
\Gamma_c(v,m_e^2,M^2;M_\gamma) \nn *
&& {} + \Gamma_1(v,m_e^2,M^2) \nn *
&& {} + \Gamma_2(v,m_e^2,M^2) .
\eeqa
The explicit expressions for 
$\Gamma_c$, $\Gamma_1$, $\Gamma_2$ are displayed in Appendix \ref{appA}. 
The function $\Gamma_c$, containing a logarithmic
dependence on the infrared regulator $M_\gamma$,  
corresponds to the long-distance 
component of the loop amplitudes which generates infrared and Coulomb 
singularities. In the case of the $K^+$ decay, the Coulomb singularity is 
outside the physical region, while it occurs on its boundary for the $K^0$ 
decay.
The other terms represent the remaining nonlocal photon loop contribution.

Note that the effective form factor $f_+(t)$ depends only on the
single variable $t$. This can be achieved \cite{CKNRT02} in the case of 
$K_{e3}$ decays by the decomposition defined by
(\ref{factor1}) and (\ref{Gamma}).  The explicit form of
$f^{K^0\pi^-}_+(t)$ and $f^{K^+\pi^0}_+(t)$ will be reviewed in the
next section.

In order to arrive at an infrared-finite (observable) result, also the 
emission of a real photon has to be taken into account.
The radiative amplitude ${\cal M}^\gamma$ can be expanded in powers of the 
photon energy $E_\gamma$, 
\beq 
{\cal M}^\gamma = {\cal M}_{(-1)}^{\gamma}
+ {\cal M}_{(0)}^{\gamma} + \dots  , 
\label{low1}
\eeq 
where 
\beq {\cal
M}_{(n)}^{\gamma} \sim (E_\gamma)^n   .  
\eeq 
Gauge invariance relates ${\cal M}_{(-1)}^{\gamma}$ 
and ${\cal M}_{(0)}^{\gamma}$ to the 
non-radiative amplitude ${\cal M}$, and thus to the full form factor 
$F_+(t,v)$. 
Upon taking the square modulus and summing over spins, the radiative
amplitude generates a correction $\rho^\gamma (y,z)$ to the Dalitz
plot density of (\ref{newrho}). The observable distribution is 
now the sum 
\beq 
\rho (y,z) =  \varrho (y,z) \, + \rho^\gamma  (y,z)   .  
\eeq
Both terms on the right hand side of this equation  
contain infrared divergences (from virtual or real soft photons).  
Upon using (\ref{factor1}) and expanding to first order in $\alpha$,            
the observable density can be written in terms of a new kinematical 
density $A_1$ \cite{CKNRT02}, and  the effective form 
factor $f_+ (t)$ defined in (\ref{factor1}), 
\beq \label{fulldensity}
\rho (y,z) = {\cal N} \,  S_{\rm EW} \,    
A_1 (y,z)  \, |f_{+} (t)|^2 ,  
\eeq
where we have pulled out the short-distance enhancement factor \cite{MS93} 
\beq
S_{\rm EW} := 
S_{\rm EW}(M_\rho,M_Z) .
\eeq

The kinematical density $A_1$ is given by \cite{CKNRT02}
\beq \label{deltas}
A_1 (y,z) = A_1^{(0)} (y,z) \,  \left[ 1 + \Delta^{\rm IR} (y,z) \right] 
\,  + 
 \, \Delta_{1}^{\rm IB} (y,z)   . 
\eeq
The function $\Delta^{\rm IR} (y,z)$ arises by combining the contributions 
from $|{\cal
M}_{(-1)}^{\gamma}|^2$ and $\Gamma (v,m_e^2,M^2; M_\gamma)$. 
Although the individual contributions contain infrared divergences, the 
sum is  finite.  
The factor $\Delta^{\rm IB}_1 (y,z)$ originates from averaging the 
remaining terms of $|{\cal M}^{\gamma}|^2$ [see (\ref{low1})]  
and are infrared-finite. Note that both $\Delta^{\rm IR} (y,z)$ and 
$\Delta^{\rm IB}_1 (y,z)$ are sensitive to the treatment 
of real photon emission in the experiment. 
A detailed analysis of these corrections for the $K^+_{e3}$ decay was 
performed in \cite{CKNRT02}. The analogous discussion for the $K^0_{e3}$ 
case will be given in Sects. \ref{sec: Real photon radiation} and 
\ref{sec: Extraction}.

Finally, integration over the Dalitz plot allows one to define 
the infrared-safe partial width, from which one extracts 
eventually the CKM element $|V_{us}|$.   
With the linear expansion of the effective form factor, 
\beq \label{ex1}
f_{+}^{K \pi} (t) = f_{+}^{K \pi} (0) \ \bigg( 1 +  
 \frac{t}{M_{\pi^{\pm}}^2} \lambda_+^{K \pi} \bigg)  , 
\eeq
the infrared-finite decay rate 
\beq 
\Gamma(K_{e 3 (\gamma)}) := 
\Gamma(K \to \pi  e^+  \nu_e) +
\Gamma(K \to \pi e^+ \nu_e \gamma) 
\eeq
can be expressed as 
\beq
\Gamma(K_{e 3 (\gamma)}) = 
{\cal N} \,   S_{\rm EW} \, 
\big| f_{+}^{K \pi} (0) \big|^2   \, I_{K}  ,
\label{ex2}
\eeq
where 
\beqa
I_{K}  &=&  \int\limits_{\cal D}  dy \, dz \, A_1 
(y,z) 
\bigg( 1 + \frac{t}{M_{\pi^{\pm}}^2} \lambda_+^{K \pi} \bigg)^2 
\nonumber \\
&=& a_0 + a_1  \,  \lambda_+^{K \pi}  + a_2 \,  
(\lambda_{+}^{K  \pi})^2  . 
\label{ex3}
\eeqa

In principle, one could easily go beyond the linear approximation 
(\ref{ex1}) for the determination of the phase space integral. Indeed, the 
curvature of the form factor, which has been neglected in (\ref{ex1}), is 
determined by (numerically unknown) coupling constants arising at 
$\cO(p^6)$ in the chiral expansion \cite{BT03}. A measurement of this 
curvature term in future experiments would be highly welcome. However, in 
view of the present experimental and theoretical situation, we restrict 
ourselves to the linear approximation (\ref{ex1}). In our analysis, we 
are using the experimentally determined values of the slope parameters. 
This method \cite{CKNRT02} minimizes the uncertainties in the 
determination 
of the phase space integrals for the time being.

In order to extract $|V_{us}|$ at the $\sim 1 \% $ level, we have to
provide a theoretical estimate of the form factor $f_{+}^{K \pi}$ at
$t=0$ and of the phase space integral in presence of isospin breaking and
electromagnetic effects. We devote the next two sections to these
tasks. 

\section{The form factors 
$\mbox{\boldmath $f_+^{K^0 \pi^-}(t)$}$
and
$\mbox{\boldmath $f_+^{K^+ \pi^0}(t)$}$
}
\label{sec: form factors}
\renewcommand{\theequation}{\arabic{section}.\arabic{equation}}
\setcounter{equation}{0}

In this section we review the structure of the $K_{e 3}$ form factors
in the framework of chiral perturbation theory, including
contributions of order $p^4$ (with isospin breaking) \cite{gl852} and
$e^2 p^2$ \cite{CKNRT02}, as well as $p^6$ effects in the isospin
limit \cite{BT03,PS02}.
 
It is convenient \cite{CKNRT02} to write the effective form factor as the 
sum of two terms,
\beq
f_+(t) = \wt{f}_+ (t) \,+\, \wh{f}_ +  .
\label{eff}
\eeq
The first one represents the pure QCD contributions  
(in principle at any order in the chiral expansion) plus 
the electromagnetic contributions up to order $e^2 p^2$ 
generated by the non-derivative  Lagrangian 
\beq
\cL_{e^2 p^0} = e^2 F_0^4 Z \langle \Q_L^{\rm em} \Q_R^{\rm em} \rangle . 
\eeq
Diagrammatically, they arise from purely mesonic 
graphs. In the definition of $\wt{f}_{+}^{K^+ \pi^0} (t)$, 
we have included also the electromagnetic counterterms relevant 
to $\pi^0$--$\eta$ mixing. The second term in (\ref{eff}) represents the 
local effects of virtual photon exchange of order $e^2 p^2$.

\subsection{Formal expressions}

The explicit form of $\wt{f}_{+}^{K^0 \pi^-} (t)$ 
is given by \cite{gl852} 
\begin{eqnarray} \label{fplusK0pims}
 \wt{f}_{+}^{K^0 \pi^-} (t)  
& = & 
1 + \frac{1}{2} H_{K^+ \pi^0} (t) +  
 \frac{3}{2} H_{K^+ \eta} (t) +  H_{K^0 \pi^-} (t)    \nonumber \\*
&&{} + \sqrt{3} \, \ve^{(2)} \left[ H_{K \pi} (t) -  
H_{K \eta} (t) \right] + \dots  ,  
\end{eqnarray}
where the ellipses indicate contributions of higher orders in the chiral 
expansion (see below for the inclusion of the  ${\cal O} (p^6)$ 
term in the isospin limit). 
The function $H_{PQ}(t)$ \cite{gl852,GL85} is reported in 
Appendix \ref{appB}. The leading order $\pi^0$--$\eta$ mixing 
angle $\ve^{(2)}$ is given by
\beq \label{eps2}
\ve^{(2)} = \frac{\sqrt{3}}{4} \frac{m_d-m_u}{m_s-\wh{m}}, \quad
\wh{m} = (m_u + m_d)/2 .
\eeq
The local electromagnetic term takes the form \cite{CKNRT02}
\beqa \label{fhK0}
\wh{f}_{+}^{K^0 \pi^-}
&=& 4\pi\alpha \Big[
2K_{12}^r(\mu) \, + \, \frac{4}{3} X_1 - \frac{1}{2} \wt{X}_6^r(\mu) 
\nn
&-& 
\frac{1}{32\pi^2} \Big( 3 + \log\frac{m_e^2}{M_{\pi^\pm}^2} + 
3\log\frac{M_{\pi^\pm}^2}{\mu^2} \Big)\Big]
.
\eeqa
The parameter $K_{12}^r(\mu)$ denotes the renormalized 
(scale dependent) part of the coupling constant $K_{12}$ introduced 
in the effective Lagrangian of order $e^2 p^2$ \cite{urech} 
describing 
the interaction of dynamical photons with hadronic degrees of freedom 
\cite{nr95,nr96}. 
The ``leptonic" couplings $X_1$, $X_6$ have been defined 
in \cite{lept}.  The coupling 
constant $\wt{X}_6^r(\mu)$ is obtained from  $X_6^r(\mu)$ after 
the subtraction of the short-distance contribution \cite{CKNRT02},
\beq \label{subtr}
X_6^r(\mu) = X_6^{\rm SD} + \wt{X}_6^r(\mu),
\eeq
where
\beq
e^2 X_6^{\rm SD} = -\frac{e^2}{4 \pi^2} \log \frac{M_Z^2}{M_\rho^2}
= 1 - S_{\rm EW}(M_\rho, M_Z),
\eeq
which defines \cite{MS93} also the short-distance enhancement factor 
$S_{\rm EW}(M_\rho, M_Z)$ to leading order. Including also leading 
QCD correction \cite{MS93}, it assumes the numerical value
\beq
S_{\rm EW} = 1.0232.
\eeq

We list here also the contributions to the 
$K^+_{e 3}$ form factor $f_{+}^{K^+ \pi^0} (t)$. Displaying only terms 
up to $\cO (p^4)$, the mesonic loop contribution is given by \cite{CKNRT02}
\begin{eqnarray}
 \wt{f}_{+}^{K^+ \pi^0} (t)  
& = & 1 + \sqrt{3} \,  \Big( \ve^{(2)} + 
\ve^{(4)}_{\rm S} + \ve^{(4)}_{\rm EM} \Big) \nonumber \\*
&&  +  \frac{1}{2} H_{K^+ \pi^0} (t) +  
 \frac{3}{2} H_{K^+ \eta} (t) +  H_{K^0 \pi^-} (t)    \nonumber \\*
&&  +  \sqrt{3} \, \ve^{(2)} \Bigg[
 \frac{5}{2} H_{K \pi} (t) + \frac{1}{2} H_{K \eta} (t) \Bigg]
+ \dots , \nn
&&
\label{ff1}
\end{eqnarray}
The pure QCD part of this expression was given in 
\cite{gl852}, the inclusion of electromagnetic contributions 
to the meson masses and the additional contribution
of $\cO(e^2 p^2)$ 
due to $\pi^0$--$\eta$ mixing 
\cite{nr95}, were added in \cite{nr96}. 
The sub-leading contributions to the $\pi^0$--$\eta$ mixing angle entering 
in (\ref{ff1}) are 
\beqa \label{eps4S}
\lefteqn{\ve^{(4)}_{\rm S} = 
- \frac{2 \, \ve^{(2)}}{3 (4 \pi F_0)^2 (M_{\eta}^2 - M_{\pi}^2)}} \nonumber 
\\* 
& \times &  \bigg\{ (4 \pi)^2 \, 64 \left[3 L_7 + 
L_8^r (\mu) \right] 
(M_K^2 - M_\pi^2)^2 
\nn
&& {} -  M_\eta^2 (M_K^2 - M_\pi^2) \log \frac{M_\eta^2}{\mu^2}
 +  M_\pi^2 (M_K^2 - 3 M_\pi^2) \log \frac{M_\pi^2}{\mu^2}  \nonumber \\
&& {} - 2 M_K^2 (M_K^2 - 2 M_\pi^2) \log \frac{M_K^2}{\mu^2} 
- 2 M_K^2 (M_K^2 - M_\pi^2) \bigg\}  , \nn
&&   
\eeqa
and 
\beqa \label{eps4EM}
\lefteqn{\ve^{(4)}_{\rm EM} =   
\frac{2 \, \sqrt{3} \, \alpha \, M_K^2}{108 \, \pi \, (M_\eta^2 
-M_\pi^2)} } \no \\*
& \times & \bigg\{ 2 (4 \pi)^2 \Big[
-6  K_3^r (\mu) + 3 K_4^r (\mu)  
+ 2 K_5^r (\mu) + 2 K_6^r (\mu) \Big]  \nonumber \\
&& {} -  9 Z \left(\log \frac{M_K^2}{\mu^2} + 1 \right) \bigg\} . 
\eeqa  
The local electromagnetic contribution for $K^+_{e 3}$ is given by 
\beqa
\wh{f}_{+}^{K^+ \pi^0} &=& 4\pi\alpha \Big[
2K_{12}^r(\mu) - \frac{8}{3} X_1 - \frac{1}{2} \wt{X}_6^r(\mu) 
\nonumber\\
&-& \frac{1}{32\pi^2} \Big(
3 + \log\frac{m_e^2}{M_{K^\pm}^2} + 3\log\frac{M_{K^\pm}^2}{\mu^2}
\Big)\Big] .
 \label{factor2}
\eeqa

What is still missing in the expressions (\ref{fplusK0pims}) and 
(\ref{ff1}), is the contribution of order $p^6$. Neglecting 
isospin breaking effects at this order, the form factors of both 
processes receive an 
equal shift which has been calculated rather recently \cite{PS02,BT03} in 
terms of  loop functions (containing some of the $L_i$) and certain 
combinations of the coupling constants $C_i$ \cite{p6,BCE00} arising at 
order $p^6$ in the chiral expansion. For our purposes, we will need only 
the value of this contribution at $t = 0$ \cite{BT03},
\beqa \label{p6contr}
 \wt{f}_{+}^{K \pi} (0)  \Big|_{p^6}  
&=& - \, 8 \,  \left( \frac{M_K^2-M_\pi^2}{F_\pi^2} 
\right)^2 \, \left[ C_{12}^r(\mu)+C_{34}^r(\mu) \right] \nonumber \\
& & \, + \ \Delta_{\rm loops} (\mu)   .
\eeqa

\subsection{Numerical estimates}

In view of the subsequent application to the extraction of $|V_{us}|$ from
$K_{e3}$ partial widths, we report here numerical estimates for the
vector form factor $f_{+}^{K \pi} $ at zero momentum transfer ($t=0$).
We recall here that in principle also the slope parameter $\lambda_{+}^{K
\pi}$ can be predicted within chiral perturbation theory. However, due
to the relatively large uncertainty induced by the low energy constant
$L_{9}^r (M_\rho)$, we shall use the measured
value of $\lambda_{+}^{K \pi}$ in the final analysis.

Apart from meson masses and decay constants, which lead to negligible
uncertainties, the vector form factor depends on a certain number of
parameters (quark mass ratios and low energy constants), whose input
we now summarize.

For the quark mass ratio $\ve^{(2)}$ defined in
(\ref{eps2}) we use \cite{Leutwyler96}
\beq \label{epsnum} 
\ve^{(2)} = (1.061 \pm 0.083) \times 10^{-2} .    
\eeq 
This number is consistent with the one obtained from a $p^6$ fit
\cite{ABT01} of the input parameters of chiral perturbation theory
within the large errors of the latter analysis. 

For the particular combination of $L_i$ entering in (\ref{eps4S}), we
take
\beq
3 L_7 + L_8^r(M_\rho) = (-0.33 \pm 0.08) \times 10^{-3},
\eeq
which is again consistent with the analysis of order $p^6$ in 
\cite{ABT01}. 

For the relevant combination of electromagnetic low energy couplings
appearing in (\ref{eps4EM}), we use \cite{BP97} 
\beqa
\wh{K}^r(M_\rho) &:=& (-6K_3+3K_4+2K_5+2K_6)^r (M_\rho) \nn
&=& (5.7 \pm 6.3) \times 10^{-3} , 
\eeqa
while for the coupling constant $K_{12}$
entering in the purely electromagnetic part (\ref{fhK0}, \ref{factor2}) 
we take \cite{moussallam}:
\beq
\label{k12}
K^r_{12} (M_\rho) = (-4.0 \pm 0.5) \times 10^{-3} . 
\eeq

Finally, for the (unknown) ``leptonic'' constants we may resort to the usual 
bounds suggested by dimensional analysis:
\beq \label{diman}
|X_1|, \ |\wt{X}^r_6(M_\rho)| \leq 1/(4 \pi)^2 \simeq 6.3 \times 10^{-3} . 
\eeq
An alternative strategy will be discussed in Sect. \ref{sec: comb}.

The above numerical input allows us to evaluate the form factor 
for both $K^0 \pi^-$ and $K^+ \pi^0$ transitions. 
To order $p^4$, the QCD part (\ref{fplusK0pims}) of the 
form factor 
at $t=0$ is uniquely determined in terms of physical meson masses (apart 
from a 
tiny contribution proportional to the leading order $\pi^0$--$\eta$ 
mixing angle):
\beq
\wt{f}_{+}^{K^0 \pi^-} (0)  = 0.97699 \pm 0.00002 . 
\eeq  

Using (\ref{k12}) and (\ref{diman}),  we find 
\beqa \label{fem0}
\wh{f}_{+}^{K^0 \pi^-} &=& 0.0046 \pm 0.0001 \pm 0.0008 \pm 0.0003
\nonumber \\
&=&  0.0046 \pm 0.0008  
\eeqa
for the local electromagnetic contribution to the form factor.
The errors given in the first line of (\ref{fem0})  correspond to the 
uncertainties 
of $K_{12}^r$, $X_1$ and $\wt{X}_6^r$. 
In this term, the relative uncertainty is almost exclusively
due to the poor present knowledge of $X_1$. Despite this, in the final
result for $f_{+}^{K^0 \pi^-} (0)$ this is an effect of only $0.08
\%$.     

Combining the values given above, we obtain the result at 
$\cO(p^4, (m_d-m_u) p^2, e^2 p^2)$: 
\beq \label{fplusnum}
f_{+}^{K^0 \pi^-} (0)  = 0.9816 \pm 0.0008  .
\eeq 

To this value, we have to add the contribution (\ref{p6contr}) of
order $p^6$, which suffers from a much larger uncertainty. 
Before turning to this issue, we also list
the corresponding results for the $K^+ \pi^0$ form factor at 
$\cO(p^4, (m_d-m_u) p^2, e^2 p^2)$ \cite{CKNRT02}:
\beqa 
\wt{f}_{+}^{K^+ \pi^0} (0)  & = & 1.0002  \pm 0.0022, \\ 
\wh{f}_{+}^{K^+  \pi^0} &=& 0.0032 \pm 0.0016, \\
f_{+}^{K^+ \pi^0} (0)  &=& 1.0034 \pm 0.0027  .
\label{fplusnum2}
\eeqa

\subsection{The $\mbox{\boldmath $p^6$}$ contribution}

Being the largest source of theoretical uncertainty in the extraction
of $|V_{us}|$, the $p^6$ contribution (\ref{p6contr}) deserves a
separate discussion. 
The loop part is given by \cite{BT03}
\beq \label{Deltaloops}
\Delta_{\rm loops} (M_\rho) = 0.0146 \pm 0.0064 . 
\eeq 
The quoted error reflects the uncertainty in the $p^4$ couplings $L_i^r$
(contributing at order $p^6$ through insertions in one-loop diagrams), 
as well as a conservative estimate of higher order effects 
\cite{BT03}. 
Concerning the local contribution in (\ref{p6contr}), 
\beq \label{p6loc}
\wt{f}_{+}^{K \pi} (0) \Big|_{p^6}^{\rm local} =
- 8 \left( \frac{M_K^2-M_\pi^2}{F_\pi^2} 
\right)^2 \left[ C_{12}^r(M_\rho)+C_{34}^r(M_\rho) \right] ,
\eeq     
there are at present several open questions.  As pointed out
in \cite{BT03} the couplings $C_{12}^r (\mu)$ and $C_{34}^r (\mu)$ are
experimentally accessible in $K_{\mu 3}$ decays, as they are related
to slope and curvature of the scalar form factor $f_0 (t)$.
Experimental efforts in this direction have started, and in the long
run this approach will give the most reliable result.  
For the time being, following \cite{BT03} we identify the estimate 
of short range contributions to $f^{K \pi}_+(0)$ given in \cite{lr84} 
with (\ref{p6loc}): 
\beq \label{p6loclr}
\wt{f}_{+}^{K \pi} (0) \Big|_{p^6}^{\rm local} = -0.016 \pm 0.008 .
\eeq  
A value of this size seems to be supported by a recent 
coupled channels dispersive analysis of the 
scalar form factor \cite{jop04}, 
and can also be obtained by resonance saturation 
\cite{CPEN03} for the couplings entering in (\ref{p6loc}), 
\beqa
C_{12}^{\rm res} &=& - \frac{F_\pi^2}{2} \frac{c_d c_m}{M_S^4} , \nn
C_{34}^{\rm res} &=& \frac{F_\pi^2}{2} \left( 
\frac{c_d c_m + c_m^2}{M_S^4} + \frac{d_m^2}{M_P^4}
\right) .
\eeqa
Using \cite{AP02}
\beq
c_m = c_d = F_\pi/2, \, d_m = F_\pi /2 \sqrt{2}, \, M_P = \sqrt{2} 
M_S,
\eeq
we obtain
\beq
C_{12}^{\rm res} = - \frac{1}{8} \left( \frac{F_\pi}{M_S} \right)^4,
\quad
C_{34}^{\rm res} = \frac{17}{64} \left( \frac{F_\pi}{M_S} \right)^4.
\label{p6lecs}
\eeq
Inserting $M_S = 1.48 \, {\rm GeV}$ (scenario A of \cite{CPEN03}), we find
\beq
\wt{f}_{+}^{K \pi} (0) \Big|_{p^6}^{\rm local} = -0.012 ,
\eeq
fully consistent with (\ref{p6loclr})\footnote{
We should remark here that the estimate (\ref{p6lecs}) is not the
complete resonance saturation result, which actually involves more
resonance couplings \cite{CPEN03}. It represents, however, a well
defined starting point and  further work along these lines  should 
provide the size of missing contributions and an estimate of 
the uncertainty}.  

It is important to stress here that the above methods do not specify
the chiral renormalization scale at which the estimate of the relevant
$p^6$ couplings applies.  This in turn leads to an intrinsic ambiguity
in the final answer, as the chosen reference scale $\mu=M_\rho=0.77$
GeV is somewhat arbitrary.  The impact of this effect can be
quantified by studying the scale dependence of $C_{12}^r + C_{34}^r$
(or equivalently of $\Delta_{\rm loops}$) with renormalization group
techniques \cite{BCE00}.  We find $\Delta_{\rm
loops}(1\mbox{GeV})= 0.0043$ and  $\Delta_{\rm loops}(M_\eta)=
0.0310$. We conclude that the present uncertainty on the $p^6$
contribution to $f_+^{K \pi}(0)$ is {\em at least} 0.01.

Keeping in mind the above caveats, as a net effect, 
there is a large destructive interference between  
the loop part (\ref{Deltaloops}) and the local contribution (\ref{p6loclr})
and we arrive at
\beq \label{p6contrib}
\wt{f}_{+}^{K \pi} (0) \Big|_{p^6} = -0.001 \pm 0.010 .
\eeq 
Adding this number to the ones in (\ref{fplusnum}) and (\ref{fplusnum2}), 
we obtain our final values at $\cO(p^6, (m_d-m_u) p^2, e^2 p^2)$:
\beqa \label{fK0fin}
f_{+}^{K^0 \pi^-} (0) &=& 0.981 \pm 0.010   , \\
f_{+}^{K^+ \pi^0}(0) &=& 1.002 \pm 0.010 .
\label{fK0fin2}
\eeqa 
We remark here that previous analyses \cite{CKNRT02,VC03} of $K_{e3}$
decays and $|V_{us}|$ did not include the $p^6$ loop contribution
$\Delta_{\rm loops} (M_\rho)$, and that further work is needed to  
clarify whether the uncertainty in (\ref{fK0fin}) and (\ref{fK0fin2}) 
is a realistic one.

\section{Real photon radiation in $\mbox{\boldmath $K^{0}_{e3}$}$}
\label{sec: Real photon radiation}
\renewcommand{\theequation}{\arabic{section}.\arabic{equation}}
\setcounter{equation}{0}

\subsection{Photon-inclusive decay distribution}

\begin{figure}[h]
\hspace*{0.0cm}
\resizebox{0.45\textwidth}{!}{
  \includegraphics{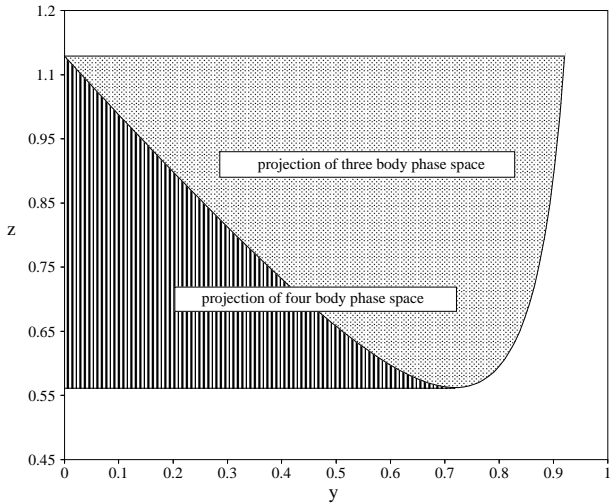}}
\caption{Dalitz plot for the three and four body final states.}
\label{dplot}    
\end{figure}

We present here in detail a possible treatment of the contribution of the 
real photon emission process
\beq
K^0_L(p_K) \to \pi^- (p_\pi) \,  e^+ (p_e) \, 
\nu_e (p_\nu) \, \gamma (p_\gamma)  \ ,  
\eeq
in complete analogy with the procedure proposed in \cite{gin67} and 
\cite{CKNRT02} for the analysis of the $K^+_{e3}$ decay. To this end we 
define the kinematical variable \cite{gin68}
\beq \label{x}
x = (p_\nu + p_\gamma)^2 = (p_K - p_\pi - p_e)^2  ,
\eeq
which determines the angle between the pion and positron 
momentum for given energies $E_\pi$, $E_e$. 
For the analysis of the experimental data, we suggest to 
accept all pion and positron energies
within the whole $K^{0}_{e3}$ Dalitz plot $\D$ given by 
\beqa \label{domain1}
2 \sqrt{r_e} & \leq y \leq & 1 + r_e - r_\pi  , \nonumber \\ 
a (y) - b (y) & \leq z \leq & a (y) + b (y)  ,  
\eeqa
where  
\beqa \label{domain2}
a (y) & = & \frac{(2 - y) \, (1 + r_e + r_\pi -y)}{2 ( 1 + r_e - y)}
 , \nonumber \\ 
b (y) & = & \frac{\sqrt{y^2 - 4 r_e} \, 
(1 + r_e - r_\pi -y)}{2 ( 1 + r_e - y)} ,   
\eeqa
or, equivalently,
\beqa \label{domain3}
2 \sqrt{r_\pi} & \leq z \leq & 1 + r_\pi - r_e  , \nonumber \\ 
c (z) - d (z) & \leq y \leq & c (z) + d (z)  ,  
\eeqa
where  
\beqa \label{domain4}
c (z) & = & \frac{(2 - z) \, (1 + r_\pi + r_e -z)}{2 ( 1 + r_\pi - z)}
 , \nonumber \\ 
d (z) & = & \frac{\sqrt{z^2 - 4 r_\pi} \, 
(1 + r_\pi - r_e -z)}{2 ( 1 + r_\pi - z)}  , 
\eeqa
and all kinematically 
allowed values of the Lorentz invariant $x$ defined in (\ref{x}). 
Note that this prescription excludes a part of the pure $K_{e3\gamma}$ events.
The situation is best explained by Figure~\ref{dplot}. The dotted area 
refers to the $K^0_{e3}$ Dalitz plot, whereas the striped region shows which
part of the projection of the $K^0_{e3\gamma}$ phase space onto the $(y,z)$ 
plane is excluded. 
This translates into the distribution 
\beqa \label{rhogamma}
\rho^\gamma (y,z) &=&  \frac{M_K}{2^{12}\pi^5} 
\int\limits_{M^2_\gamma}^{x_{\rm max}} 
dx 
\frac{1}{2 \pi}  \int  \frac{d^3 p_\nu}{p_\nu^0} 
\frac{d^3 p_\gamma}{p_\gamma^0} \nn 
&\times& \delta^{(4)}(p_K-p_\pi-p_e-p_\nu-p_\gamma)  
\sum_{\rm pol}  |{\cal M}^{\gamma}|^2   , \nn
&& 
\eeqa 
with
\beqa \label{xmax} 
x_{\rm max} &=& M_K^2 
\bigg\{1 + r_\pi + r_e - y - z  \nn
&&{}+ \frac{1}{2} \Big[y z  +  \sqrt{(y^2 - 4 r_e)(z^2 - 4 r_\pi)} 
\Big] \bigg\} . 
\eeqa
In (\ref{rhogamma}) we have extended the integration over the whole range 
of the invariant mass of the unobserved $\nu_e \, \gamma$ system.
The integrals occurring in (\ref{rhogamma}) have the general form 
\cite{gin67}
\beqa \label{Imn}
\lefteqn{I_{m,n}(p_1,p_2;P,M_{\gamma}) :=} \nonumber \\* 
&& \frac{1}{2 \pi} \int 
\frac{d^3 q}{q^0} \frac{d^3 k}{k^0} 
\frac{\delta^{(4)}(P-q-k)} 
{(p_1 \! \cdot k \! 
+M_{\gamma}^2/2)^m (p_2 \! \cdot \! k + M_{\gamma}^2/2)^n}  . \nonumber 
\\*
&&
\eeqa
The results for these integrals in the limit $M_\gamma = 0$ can be found 
in the Appendix of \cite{gin67}.
Using the definition (\ref{Imn}), the  radiative decay 
distribution (\ref{rhogamma}) can be written as 
\cite{gin68}
\beqa \label{decomp}
\lefteqn{\rho^\gamma (y,z) =} \nonumber \\* 
&& \frac{\alpha}{\pi} \Bigg[ \rho^{(0)} (y,z) I_0(y,z;M_\gamma) 
+\frac{G_{\rm F}^2 |V_{us}|^2 |f_{+}^{K^0 \pi^-}|^2 M_K}{64 \pi^3} \nn
&& \times \! \! \int\limits_0^{x_{\rm max}} \! \! dx 
\sum_{m,n} c_{m,n} I_{m,n} \Bigg]  ,
\nonumber \\*
&&
\eeqa
where the infrared divergences are now confined to\footnote{The right-hand 
side of the corresponding expression (6.7) in \cite{CKNRT02} should be 
multiplied by $1/4$}
\beqa \label{I0}
\lefteqn{I_0(y,z;M_\gamma) =}  \nn
&& \frac{1}{4} \int\limits_{M^2_{\gamma}}^{x_{\rm max}} dx 
\Big[ 2 \, p_e \! \cdot \! p_{\pi} \, 
I_{1,1}(p_e,p_\pi;p_K-p_\pi-p_e;M_\gamma) 
\nn[-14pt]
&& \qquad \quad {} - M_{\pi}^2 \,
I_{0,2}(p_e,p_\pi;p_K-p_\pi-p_e;M_\gamma)
\nn
&& \qquad \quad {}- m_{e}^2 \,
I_{2,0}(p_e,p_{\pi};p_K-p_\pi-p_e;M_\gamma) \Big] . 
\eeqa
The explicit form of the function $I_0$ can be found in 
Appendix \ref{appC}.
The coefficients $c_{m,n}$ were given in Eq. (21) of \cite{gin68}.

\begin{table*}
\caption{\label{table2} $A_1^{(0)}(y,z) \times 10^2 $ for $K^0_{e3}$ decay
}
\vspace{0.2cm}
\begin{center}
\begin{tabular}{|c|c|c|c|c|c|c|c|c|c|}
\hline
$z \backslash  y$& 0.05 &0.15 &0.25&0.35&0.45&0.55&0.65&0.75&0.85 
\\
\hline
1.05 & 6.54 & 36.54 & 58.54 & 72.54 & 78.54 & 76.54 & 66.54 &48.54 & 22.54
\\
\hline
1.00 &  & 19.54 & 43.54 & 59.54 & 67.54 & 67.54 & 59.54 &43.54 & 19.54   
\\
\hline
0.95&  & 2.54 & 28.54 & 46.54 & 56.54 & 58.54 & 52.54 & 38.54 & 16.54 
\\
\hline
0.90& & & 13.54 & 33.54 & 45.54 & 49.54 & 45.54 & 33.54 &13.54 
\\
\hline
0.85& & & & 20.54 & 34.54 & 40.54 & 38.54 & 28.54 & 10.54   
\\
\hline
0.80& & & & 7.54 & 23.54 & 31.54 & 31.54 & 23.54 & 7.54  
\\
\hline
0.75& & & & & 12.54 & 22.54 & 24.54 & 18.54 & 4.54  
\\
\hline
0.70& & & & &  1.54 & 13.54 & 17.54 & 13.54 & 1.54 
\\
\hline
0.65& & & & & & 4.54 & 10.54 & 8.54 & 
\\
\hline
0.60& & & & & & & 3.54  & 3.54 &  
\\
\hline
\end{tabular}
\end{center}
\end{table*}

\begin{table*}
\caption{\label{table3} $[A_1 (y,z) - A_1^{(0)}(y,z)] 
\times 10^2 $ for $K^0_{e3}$ decay
}
\vspace{0.2cm}
\begin{center}
\begin{tabular}{|c|c|c|c|c|c|c|c|c|c|}
\hline
$z \backslash  y$& 0.05 &0.15 &0.25&0.35&0.45&0.55&0.65&0.75&0.85 
\\
\hline
1.05 & 1.685 & 2.071 & 1.513 & 0.562 &-0.523 &-1.541 & -2.295 & -2.542& -1.864

\\
\hline
1.00 &  & 2.188 & 1.999 & 1.243 & 0.237 &-0.799 & -1.651 & -2.061 &-1.582    
\\
\hline
0.95&  & 1.775 & 2.024 & 1.464 & 0.562 & -0.432 & -1.295 & -1.757 & -1.356 
\\
\hline
0.90& & & 1.844 & 1.524 & 0.749 & -0.180 & -1.022 & -1.501 & -1.140 
\\
\hline
0.85& & & & 1.476 & 0.856 & 0.012 & -0.792 & -1.269 & -0.925   
\\
\hline
0.80& & & & 1.313 & 0.901 & 0.162 & -0.589 & -1.049 & -0.705 
\\
\hline
0.75& & & & & 0.883 & 0.276 & -0.407 & -0.839 & -0.471  
\\
\hline
0.70& & & & &  0.772 & 0.353 & -0.243 & -0.633 & -0.200 
\\
\hline
0.65& & & & & & 0.384 & -0.097 & -0.428 & 
\\
\hline
0.60& & & & & & & 0.031  & -0.212 &  
\\
\hline
\end{tabular}
\end{center}
\end{table*}

The function $\Delta^{\rm IR}$ introduced in (\ref{deltas}) can now be 
related to $I_0$ by
\beq
\Delta^{\rm IR}(y,z) = \frac{\alpha}{\pi} \left[ I_0(y,z;M_\gamma)
+ \frac{1}{2} \Gamma(s,m^2_e,M_\pi^2;M_\gamma) \right] \ . 
\label{deltaIR}
\eeq

An analytic expression of the integral occurring in the 
last line of (\ref{decomp}) was given in 
Appendix B of \cite{gin70} in terms of the quantities $V_i$:
\beq
\int\limits_0^{x_{\rm max}} dx \: \sum_{m,n} c_{m,n} I_{m,n} = 
\sum_{i=0}^7 V_i
 .
\eeq
As already noticed in \cite{CKNRT02}, the quantity $J_9(i)$ 
given in Eq. (A9) of \cite{gin70} (which 
is needed for the evaluation of $V_7=U_7$) contains two mistakes:
the plus-sign in the last line of (A9) should be replaced by a minus-sign, 
and $|\beta_i^{\rm max}|$ at the end of the first line of (A9) should 
simply read $\beta_i^{\rm max}$. The function $\Delta_i^{\rm IB}$ introduced 
in (\ref{deltas}) can now be 
written as\footnote{Setting $\xi=0$ in the expressions of \cite{gin70} 
amounts to neglect the form factor $f_{-}(t)$, which is  
an excellent approximation in $K_{e3}$ modes}
\beq \label{DeltaIB}
\Delta_1^{\rm IB} = \frac{2 \alpha}{\pi M_K^4} \sum_{i=0}^7 V_i 
\Big|_{\xi=0}  .
\eeq

The expressions in (\ref{deltaIR}) and (\ref{DeltaIB}) fully determine
the radiatively corrected decay density $A_1 (y,z)$ (\ref{deltas}). In
order to appreciate the effect of these universal long-distance
corrections, we report the kinematical density $A_1^{(0)}$ in the
absence of electromagnetism for several individual points of the
Dalitz plot in Table \ref{table2}, while the corresponding radiative
corrections entering in (\ref{deltas}) are displayed in Table
\ref{table3}. Note that the relative size of the electromagnetic
corrections for some points (especially near the boundary) exceeds the
average shift considerably.  For completeness, we display a sample of
numerical values for the kinematical densities (\ref{kin0}) and
(\ref{deltas}) also for the $K^{+}_{e3}$ decay mode in Tables
\ref{table4} and \ref{table5}.

\begin{table*}
\caption{\label{table4} $A_1^{(0)}(y,z) \times 10^2 $ for $K^+_{e3}$ decay
}
\vspace{0.2cm}
\begin{center}
\begin{tabular}{|c|c|c|c|c|c|c|c|c|c|}
\hline
$z \backslash  y$& 0.05 &0.15 &0.25&0.35&0.45&0.55&0.65&0.75&0.85 
\\
\hline
1.05 & 8.10 & 38.10 & 60.10 & 74.10 & 80.10 & 78.10 & 68.10 &50.10 & 24.10
\\
\hline
1.00 &  & 21.10 & 45.10 & 61.10 & 69.10 & 69.10 & 61.10 &45.10 & 21.10   
\\
\hline
0.95&  & 4.10 & 30.10 & 48.10 & 58.10 & 60.10 & 54.10 & 40.10 & 18.10 
\\
\hline
0.90& & & 15.10 & 35.10 & 47.10 & 51.10 & 47.10 & 35.10 &15.10 
\\
\hline
0.85& & &0.10 & 22.10 & 36.10 & 42.10 & 40.10 & 30.10 & 12.10   
\\
\hline
0.80& & & & 9.10 & 25.10 & 33.10 & 33.10 & 25.10 & 9.10  
\\
\hline
0.75& & & & & 14.10 & 24.10 & 26.10 & 20.10 & 6.10  
\\
\hline
0.70& & & & &  3.10 & 15.10 & 19.10 & 15.10 & 3.10 
\\
\hline
0.65& & & & & & 6.10 & 12.10 & 10.10 & 0.10
\\
\hline
0.60& & & & & & & 5.10  & 5.10 &  
\\
\hline
0.55&        &       &       &       &       &       &       & 0.10 &
\\  
\hline
\end{tabular}
\end{center}
\end{table*}

\begin{table*}
\caption{\label{table5} $[A_1 (y,z) - A_1^{(0)}(y,z)] \times 10^2 $ 
for $K^+_{e3}$ decay
}
\vspace{0.2cm}
\begin{center}
\begin{tabular}{|c|c|c|c|c|c|c|c|c|c|}
\hline
$z \backslash  y$ & 
	0.05 &  0.15 & 0.25  & 0.35  & 0.45  & 0.55  & 0.65  & 0.75 & 0.85 
\\
\hline
1.05 & 1.494 & 1.697 & 1.174 & 0.313 &-0.670 &-1.593 &-2.275 &-2.486 
&-1.841
\\
\hline
1.00 &       & 1.708 & 1.364 & 0.610 &-0.320 &-1.236 &-1.946 &-2.213 &-1.638   
\\
\hline
0.95&        & 1.558 & 1.378 & 0.732 &-0.128 &-1.006 &-1.704 &-1.983 
&-1.440 
\\
\hline
0.90&        &       & 1.356 & 0.821 & 0.036 &-0.796 &-1.474 &-1.758 &-1.240 
\\
\hline
0.85&        &       & 1.321 & 0.898 & 0.190 &-0.593 &-1.248 &-1.533 &-1.035   
\\
\hline
0.80&        &       &       & 0.971 & 0.341 &-0.392 &-1.021 &-1.305 
&-0.822 
\\
\hline
0.75&        &       &       &       & 0.490 &-0.191 &-0.794 &-1.075 &-0.597  
\\
\hline
0.70&        &       &       &       & 0.639 & 0.010 &-0.566 &-0.841 
&-0.348 
\\
\hline
0.65&        &       &       &       &       & 0.214 &-0.333 &-0.598 
&-0.020
\\
\hline
0.60&        &       &       &       &       &       &-0.094 &-0.340 &  
\\
\hline
0.55&        &       &       &       &       &       &       &-0.014 &  
\\
\hline
\end{tabular}
\end{center}
\end{table*}

\subsection{Phase space integrals}

Once the function $A_1 (y,z)$ is known, the numerical coefficients
$a_{0,1,2}$ entering in the phase space integral (\ref{ex3}) can be
calculated by integration over the Dalitz plot. These are 
reported in Table \ref{table1} for the $K^0_{e3}$ mode, while the  
corresponding results for $K^+_{e3}$ can be found in \cite{CKNRT02}.  
We recall once again that these numbers correspond 
to the specific prescription for the treatment of real photons described 
in the previous section: accept all pion and positron energies
within the whole $K_{e3}$ Dalitz plot $\D$ and all kinematically 
allowed values of the Lorentz invariant $x$ defined in (\ref{x}). 

A full evaluation of the phase space factor $I_{K}$
(\ref{ex3}) requires knowledge of the slope parameter.  For both modes
we employ the measured values \cite{PDG02}\footnote{For the $K^+_{e3}$ 
mode the slope parameter given in \cite{PDG02} has 
received a small change compared to the PDG 2000 number used in 
\cite{CKNRT02}, which amounts to a negligible 
difference in the final result 
},
\beqa
\lambda_+^{K^0 \pi^-} &=& 0.0291 \pm 0.0018  , \\
\lambda_+^{K^+ \pi^0} &=& 0.0278 \pm 0.0019  .
\eeqa 

For $K^0_{e3}$ decays  the final numbers 
\beqa
\left. I_{K^0} \right|_{\alpha = 0} &=& 0.10372  , \\ 
I_{K^0} &=& 0.10339 \pm 0.00063 , 
\eeqa 
reveal that radiative corrections effectively induce a negative shift
of $0.32 \%$ in the factor $I_{K^0}$.

On the other hand, for $K^+_{e3}$ one finds
\beqa
I_{K^+}|_{\alpha =0} &=& 0.10616 , \\
I_{K^+} &=& 0.10482 \pm 0.00067  ,
\eeqa
corresponding to a negative shift of $1.27 \%$ induced by the 
radiative corrections. This is essentially unchanged from the 
analysis in \cite{CKNRT02}.

\begin{table}
\caption{\label{table1} Coefficients of the $K^0_{e3}$ phase space 
integral}
\vspace{0.2cm}
\begin{center}
\begin{tabular}{|l|c|c|c|}
\hline
 & $a_0$ & $a_1$ & $a_2$ \\ \hline 
$\alpha = 0$  & $0.09390  $ & $0.3245 $ & $0.4485$ \\ \hline 
$\alpha \neq 0$  & $0.09358  $ & $0.3241 $ & $0.4475$ \\ \hline 
\end{tabular}
\end{center}
\end{table}

\section{Extraction of $\mbox{\boldmath$|V_{us}|$}$ from  
$\mbox{\boldmath  $K^{0}_{e3}$}$ decays
}
\label{sec: Extraction}
\renewcommand{\theequation}{\arabic{section}.\arabic{equation}}
\setcounter{equation}{0}

The CKM matrix element $|V_{us}|$ can be extracted 
from the 
$K^{0}_{e3}$ decay parameters by
\beq \label{Vus}
|V_{us}| = \left[ 
\frac{128 \, \pi^{3}   \, 
\Gamma(K^0_{e 3 (\gamma)})}
{G_{\rm F}^2 \,  M_{K^{0}}^{5}  \, S_{\rm EW}  \,
I_{K^0} } 
\right]^{1/2} \cdot \frac{1}{
f_+^{K^0 \pi^-}(0) }
\eeq

In spite of the unsatisfactory present status of the  $K^0_{e3}$ 
data, we 
use them here as an illustration of the application of the above formula.
\begin{itemize}
\item[$\bullet$]  With \cite{PDG02}
\beq \label{fit}
\Gamma(K^0_{e3(\gamma)})_{\rm fit} = (7.50 \pm 0.08) 
\times 10^6 \, {\rm s^{-1}}
\eeq
and (\ref{fK0fin}), we find
\beqa
|V_{us}| &=& 0.2153 \pm 0.0011 \pm 0.0007 \pm 0.0022 \nn
         &=& 0.2153 \pm 0.0026 ,
\eeqa 
where the errors correspond to
\beqa
\Delta |V_{us}| &=& |V_{us}| \left( \pm \frac{1}{2} \frac{\Delta
\Gamma}{\Gamma} \pm
0.05 \cdot \frac{\Delta \lambda_+}{\lambda_+} \pm
\frac{\Delta f_{+}(0)}{f_{+}(0)} \right) \nn
&=& |V_{us}| \big( \pm 0.5 \% \pm 0.3 \% \pm 1.0 \% \big) .
\eeqa

\item[$\bullet$] A more realistic estimate 
of the present $K^0_{e3}$ uncertainty
is most probably given by \cite{PDG02}
\beq \label{average}
\Gamma(K^0_{e3(\gamma)})_{\rm average} = (7.7 \pm 0.5) 
\times 10^6 \, {\rm s^{-1}}, 
\eeq
which implies
\beqa
|V_{us}| &=& 0.2182 \pm 0.0071 \pm 0.0007 \pm 0.0022 \nn
         &=& 0.2182 \pm 0.0075 ,
\eeqa 
corresponding to
\beq
\Delta |V_{us}| = 
|V_{us}| \big( \pm 3.3 \% \pm 0.3 \% \pm 1.0 \% \big) .
\eeq

\item[$\bullet$]  Finally, 
combining the $K^0_L$ lifetime from the PDG with the 
preliminary photon-inclusive branching ratio from KLOE \cite{KLOE03} 
${\rm BR} (K^L_{e3(\gamma)}) = 0.384 \pm 0.002_{\rm stat.}$, we find 
\footnote{The systematic uncertainty in the KLOE result is not yet known  
\cite{KLOE03}}
\beq \label{kloek0e3} 
\Gamma(K^0_{e3(\gamma)})_{\rm KLOE(prel.)} = (7.43 \pm 0.07) 
\times 10^6 \, {\rm s^{-1}}, 
\eeq
corresponding to 
\beqa
|V_{us}| &=& 0.2143 \pm 0.0010 \pm 0.0007 \pm 0.0022 \nn
         &=& 0.2143 \pm 0.0025 .
\eeqa 
\end{itemize}

Since the present statistical precision is comparable
to the one of the PDG fit, we expect that the experimental side of the
problem will improve considerably as soon as final results from 
KLOE \cite{KLOE03} and NA48 \cite{NA4803} will become available.

\section{Extraction of $\mbox{\boldmath$|V_{us}|$}$ from
$\mbox{\boldmath  $K^{+}_{e3}$}$ decays
}
\label{sec: E865}
\renewcommand{\theequation}{\arabic{section}.\arabic{equation}}
\setcounter{equation}{0}

In this section, we update our previous analysis of the 
$K^+_{e3}$ decay \cite{CKNRT02} in view of the 
new value (\ref{p6contrib}) for the contribution of order $p^6$ and the 
recent E865 result. All other parameters of the $K^+_{e3}$ analysis in 
\cite{CKNRT02} remain essentially unchanged. 
Due to the inconsistency between PDG 2002 and E865 results, we 
analyze them separately.

\begin{itemize}

\item[$\bullet$] Using the PDG-fit\footnote{For $K^+_{e3}$ the   
difference between ``fit'' and ``average'' is not 
sizeable} input 
\beq \label{Kppdg}
\Gamma(K^+_{e3(\gamma)}) = (3.93 \pm 0.05) \times 10^6 \, {\rm s}^{-1}, 
\eeq
and assuming that this number refers to the inclusive width of 
Section \ref{sec: Real photon radiation} one obtains 
\beqa
|V_{us}| &=& 0.2186 \pm 0.0014 \pm 0.0007 \pm 0.0023 \nn
         &=& 0.2186 \pm 0.0027 .
\eeqa 

\item[$\bullet$]
The $K^+_{e3(\gamma)}$ branching ratio measured by the E865
Collaboration \cite{E865}, when combined with the $K^\pm$ lifetime 
from the PDG, leads to the decay width
\beq \label{E865res}
\Gamma(K^+_{e3(\gamma)}) = (4.12 \pm 0.08) \times 10^6 \, {\rm s}^{-1}.
\eeq
Note that the value ${\rm BR}(K^+_{e3(\gamma)})$ given in \cite{E865} 
contains also events outside the $K^+_{e3}$ Dalitz plot boundary. This 
additional $0.5 \%$ contribution has been subtracted in (\ref{E865res}) 
in accordance with our prescription of the treatment of real photons.
Finally, we find
\beqa
|V_{us}| &=& 0.2238 \pm 0.0022 \pm 0.0007 \pm 0.0023 \nn
         &=& 0.2238 \pm 0.0033 .
\eeqa 
Together with $|V_{ud}|$ and $|V_{ub}|$ as shown in (\ref{CKM}), this number 
implies 
\beq 
|V_{ud}|^2 + |V_{us}|^2 + |V_{ub}|^2 - 1 = -0.0024 \pm 0.0021,
\eeq
in rather good agreement with a unitary mixing matrix. 

\end{itemize}

The sizeable disagreement between the result of E865
and the PDG-fit (from old experiments) calls for further experimental 
efforts in this decay channel.

\section{Combined analysis of $\mbox{\boldmath$K^0_{e3}$}$ and  
$\mbox{\boldmath $K^{+}_{e3}$}$ data}
\label{sec: comb}
\renewcommand{\theequation}{\arabic{section}.\arabic{equation}}
\setcounter{equation}{0}

$K^+_{e3}$ and $K^0_{e3}$ branching fractions allow for two
independent determinations of $f_{+}^{K^0 \pi^-} (0) \cdot | V_{us}|$, 
provided one brings under theoretical control isospin breaking in
the ratio of form factors at $t=0$,
\beq \label{Rdef}
r_{+0} := f_+^{K^+ \pi^0}(0) \Big/ f_+^{K^0 \pi^-}(0).
\eeq
The standard model allows a remarkably precise prediction of this 
quantity.
The contributions of
order $p^6$ as well as the couplings $X_6$ and $K_{12}$ cancel and we are
left with the expression 
\beqa \label{Rtheor} 
r_{+0}^{\rm th} &=& 
 1+\sqrt{3} \left( \ve^{(2)} 
+ \ve^{(4)}_{\rm S} + \ve^{(4)}_{\rm EM} \right) \nn
&&{} - \frac{\alpha}{4 \pi}
\log \frac{M^2_{K^{\pm}}}{M^2_{\pi^{\pm}}} 
-16 \pi \alpha X_1 + \dots \nn 
&& {} = 1.022 \pm 0.003  
-16 \pi \alpha X_1 ,
\eeqa
where the ellipses in the second line stand for 
isospin violating corrections arising at 
$\cO((m_d-m_u)p^4, e^2 p^4)$ in the chiral expansion. We expect them to 
shift the result at most by $10^{-3}$. Also these not yet determined 
contributions 
have been accounted for in the error given in the last line of 
(\ref{Rtheor}). 
Although no theoretical estimate of the coupling $X_1$ is  
presently available, there is no reason why this low energy     
constant should lie outside the range suggested by naive  
dimensional analysis (\ref{diman}).   
 Already such a rough estimate of $X_1$  shows that $r_{+0}$ is confined 
to the rather narrow band
\beq \label{band}
1.017  \leq r_{+0}^{\rm th} \leq 1.027 .
\eeq
We emphasize that sizeable deviations from this predicted range 
could only be understood as (i) failure of naive dimensional analysis for 
$X_1$ (and a dramatic one) or (ii) failure of chiral power counting.

\begin{figure}[h]
\begin{center}
\leavevmode
\begin{picture}(100,160)
\put(20,65){\makebox(50,50){\includegraphics[height= 2 in]{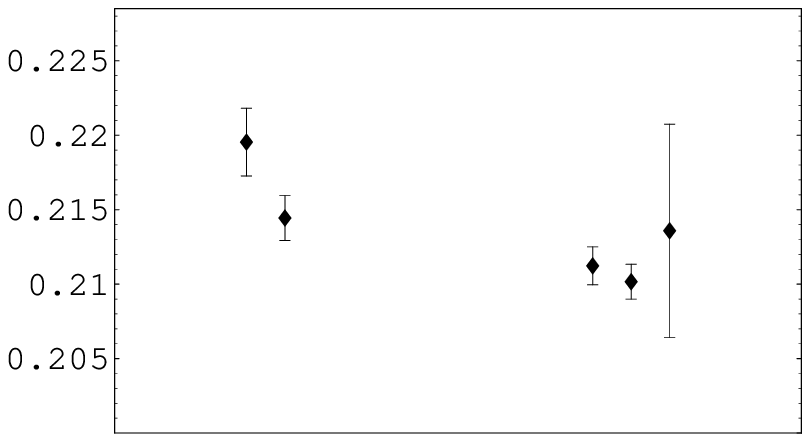}}}
\put(-75,160){\scriptsize{  $|V_{us}| \cdot f_{+}^{K^0 \pi^-} (0)$  }}
\put(0,10){
$K^+_{e3}$
}
\put(-20,130){\scriptsize{E865}}
\put(5,75){\scriptsize{PDG-FIT}}
\put(75,85){\scriptsize{PDG-FIT}}
\put(95,55){\scriptsize{KLOE$^*$}}
\put(120,120){\scriptsize{PDG-AV}}
\put(105,10){
$K^0_{e3}$
}
\end{picture}
\caption{ $|V_{us}| \cdot f_{+}^{K^0 \pi^-} (0)$ from 
$K_{e 3}$ modes (see text for details). The KLOE result is 
preliminary and the quoted error is statistical only \cite{KLOE03}. }
\label{fig:fvus}
\end{center}
\end{figure}

On the other hand, the ratio (\ref{Rdef}) is related to the observable
\beq \label{Rexp} 
r_{+0}^{\rm exp} = \left( \frac{2 \, \Gamma(K^+_{e 3
(\gamma)}) \, M_{K^0}^5 \, I_{K^0}}{\Gamma(K^0_{e 3 (\gamma)}) \,
M_{K^+}^5 \, I_{K^+}} \right)^{1/2} , 
\eeq
with the caveat that the phase space factors $I_{K}$ be evaluated 
according to the same prescription for real photons adopted 
in measuring $\Gamma(K_{e 3 (\gamma)})$.
Once again, it is instructive to consider several cases:

\begin{itemize}
\item[$\bullet$]   Using (\ref{fit}) and (\ref{E865res}), we find 
\beqa
r_{+0}^{\rm exp} &=& 1.062 \pm 0.010  \pm 0.006  \pm 0.003 
\pm 0.003  
\nn
&=& 1.062 \pm 0.013 ,
\eeqa 
where the errors given in the first line refer to the 
experimental uncertainties of $\Gamma(K^+_{e3(\gamma)})$, 
$\Gamma(K^0_{e3(\gamma)})$, $\lambda_+^{K^+ \pi^0}$ and 
$\lambda_+^{K^0 \pi^-}$, respectively. The outcome is
clearly in conflict with the prediction (\ref{band}) of the 
standard model and indicates 
indeed an inconsistency of the present $K^+_{e3}$ and    
$K^0_{e3}$ data.   
This is also illustrated by Fig. \ref{fig:fvus} where data data from
$K^+_{e3}$ (E865) and $K^0_{e3}$ (PDG-fit), after using $r_{+0}^{\rm
th}$ as discussed above, lead to two inconsistent determinations of 
the product  $f_{+}^{K^0 \pi^-} (0)\cdot | V_{us} |$.
\item[$\bullet$]  
Taking (\ref{average}) instead of (\ref{fit}), the resulting numbers  are 
\beqa
r_{+0}^{\rm exp} &=& 1.049 \pm 0.010  \pm 0.034  
 \pm 0.003  \pm 0.003 
\nn
         &=& 1.049 \pm 0.036 .
\eeqa 
This value is consistent with (\ref{band}) however 
with a large error caused by the big uncertainty in (\ref{average}). 

\item[$\bullet$]  The inconsistency is somehow mitigated when one uses 
the present PDG-fit entries for both $K^+_{e3}$ and $K^0_{e3}$, leading to 
\beqa
r_{+0}^{\rm exp} &=& 1.038 \pm 0.006  \pm 0.006  
 \pm 0.003  \pm 0.003 
\nn
         &=& 1.038 \pm 0.010 .
\eeqa 

\end{itemize}
The present confusing status is summarized in Figure \ref{fig:fvus},
where we plot $f_{+}^{K^0 \pi^-} (0)\cdot | V_{us} |$ as determined
from different  $K^+_{e3}$ and $K^0_{e3}$ experimental input\footnote{Plots 
of this type were first used in \cite{calc01} and 
can be found also in \cite{VC03,KLOE03,GIckm03}}. 
The points corresponding to $K^+_{e3}$ have been obtained by 
using the central value  for $r_{+0}^{\rm th}$.
The overall $0.78 \% $ normalization uncertainty of these points 
is not reported in the plot. 

For the analysis of forthcoming high-precision data on 
$K_{e3}$ decays we propose the following strategy: 

\begin{itemize}

\item[(a)] 
Check the consistency of $K^+_{e3}$ and $K^0_{e3}$ data by comparing 
$r_{+0}^{\rm exp}$ with the theoretically allowed range (\ref{band}).

\item[(b)] 
Determine the low energy constant $X_1$  
from $r_{+0}^{\rm exp}$  by inverting (\ref{Rtheor}),
\beq
X_1 = \frac{1.022 \pm 0.003 ({\rm theor.}) - r_{+0}^{\rm exp}}{16 \pi 
\alpha}.
\eeq 
(We refrain from extracting a number for $X_1$ based on the present data 
as they are apparently inconsistent.)

\item[(c)]
Recalculate $\wh{f}_+^{K^0 \pi^-}$ from (\ref{fhK0}) by using the 
experimentally determined parameter $X_1$.

\item[(d)]
Use the new number for $f_+^{K^0 \pi^-}(0)$ in the determination of 
$|V_{us}|$ as described in Sect. \ref{sec: Extraction}.

\item[(e)] Finally, one can also use the experimentally determined
$X_1$ to improve radiative corrections to the pion beta decay
\cite{pibeta}, relevant for the extraction of $| V_{ud} |$ from this mode
once the PIBETA experiment finalizes the analysis \cite{pibetaexp}. 

\end{itemize}

\section{Conclusions}
\label{sec: Conclusions}
\renewcommand{\theequation}{\arabic{section}.\arabic{equation}}
\setcounter{equation}{0}

In this work, we have studied $K_{e3}$ decays using chiral 
perturbation theory with virtual photons and leptons. This method allows a 
unified and consistent treatment of strong and electromagnetic 
contributions to the decay amplitudes within the standard model.  We have 
considered strong effects up to $\cO(p^6)$ in the chiral expansion. 
Isospin breaking 
due to the mass difference of the light quarks has been included up to 
the order $(m_d -m_u) p^2$. Electromagnetic effects  were taken into 
account up to 
$\cO(e^2 p^2)$. The largest theoretical error is generated by the 
contribution of $\cO(p^6)$ inducing a $1 \%$ uncertainty in the 
determination of the $K_{e3}$ form factors.  
Additional theoretical investigation is needed to increase our
confidence in the estimate of local contributions at $\cO(p^6)$.  

Based on our theoretical results, we have described the extraction of the 
CKM matrix element $|V_{us}|$ from experimental decay parameters and a 
consistency check of $K_{e3}^+$ and $K_{e3}^0$ data.

Using the recent E865 result on the $K_{e3}^+$ branching ratio, we find 
\beq
|V_{us}| =  0.2238 \pm 0.0033 ,
\eeq 
being perfectly consistent with CKM unitarity. It should be noted, 
however, that the E865 ratio differs from older $K_{e3}^+$ measurements by 
2.3 $\sigma$. Furthermore, the E865 result and the present  $K_{e3}^0$ 
rate 
as given by PDG 2002 (based on very old data) and by KLOE 
preliminary results  can hardly be reconciled  
within the framework of the 
standard model. Recently-completed or ongoing experiments will help to 
clarify the situation. 

Finally a short remark on $|V_{ud}|$, the second important source of  
information for the check of CKM unitarity: the present number for 
$|V_{ud}|$ is extracted from super-allowed Fermi transitions and neutron 
beta decay. In principle, the pionic beta decay ($\pi_{e3}$) provides 
a unique test of these existing determinations. This decay 
mode is theoretically 
extremely clean \cite{pibeta} and also completely consistent with the 
present analysis of $K_{e3}$ decays. Using the present  
result on the $\pi_{e3}$ branching ratio  from the PIBETA 
experiment \cite{pibetaexp}, one finds
\beq
|V_{ud}| =  0.9716 \pm 0.0039 ,
\eeq 
to be compared with the current PDG value shown in (\ref{CKM}). 
The  final result  from this experiment  is expected to reach a  
precision for the pion beta decay rate of about $0.5 \%$. Further efforts 
for an improvement of the experimental accuracy of $\pi_{e3}$
would be highly desirable.

\medskip

\noindent
{\small {\it Acknowledgements.} We thank J. Bijnens, G. Ecker, 
J. Gasser and H. Leutwyler for useful remarks. 
We have profited from discussions with  G. Isidori, B. Sciascia, 
T. Spadaro and J. Thompson.  
V. C. is supported by a Sherman Fairchild fellowship from Caltech.}

\section*{Appendix}

\appendix

\section{Photonic Loop Functions}
\label{appA}
\renewcommand{\theequation}{\Alph{section}.\arabic{equation}}
\setcounter{equation}{0}

The photonic loop contributions to the $K_{\ell 3}$ 
form factors depend on the charged lepton and meson masses 
$m_\ell^2$, $M^2$, as well as on the Mandelstam variables 
$u = (p_K - p_\ell)^2$ (for $K^{+}_{\ell 3}$ decays) and  
$s = (p_\pi + p_\ell)^2$ (for $K^{0}_{\ell 3}$ decays).
In what follows we denote by $v$ the Mandelstam variable 
appropriate to each decay.  
In order to express the loop functions in a compact way, 
it is useful to define the following intermediate variables:
\begin{equation}
 R  =  \frac{m_{\ell}^2}{M^2}  , \ \ 
 Y = 1 + R - \frac{v}{M^2}  , \ \     
 X = \frac{Y - \sqrt{Y^2 - 4 R}}{2 \sqrt{R}} . 
\end{equation} 
In terms of such variables and of the dilogarithm 
\begin{equation}
{\rm Li}_2 (x) = - \int_{0}^{1} \frac{dt}{t} \log (1 - x t)   , 
\end{equation}
the functions contributing to $\Gamma (v,m_{\ell}^2,M^2 ; M_\gamma)$
are given by \cite{CKNRT02}
\beqa \label{Gammac}
\lefteqn{\Gamma_c (v,m_\ell^2,M^2;M_\gamma)  =  
 2 M^2 Y \, {\cal C} (v,m_\ell^2,M^2)} \nn *  
&& {} + 2  \log \frac{M m_\ell}{M_\gamma^2} 
\bigg(1 + \frac{X Y 
\log X}{\sqrt{R} 
(1 - X^2)}  \bigg)  . 
\eeqa
\beqa
\lefteqn{{\cal C} (v,m_\ell^2,M^2)  =   \frac{1}{m_\ell M} \frac{X}{1 - 
X^2}} \nonumber \\*
&\times&
\left[  - \frac{1}{2} \log^2 X + 2 \log X \log (1 - X^2) - 
\frac{\pi^2}{6} + \frac{1}{8} \log^2 R \right. \nn
&&{}+ \left.  {\rm Li}_2 (X^2) + {\rm Li}_2  (1 - \frac{X}{\sqrt{R}}) + 
{\rm Li}_2 (1 - X \sqrt{R})  \right]  ,  \nonumber \\*
&&
\eeqa 
\beqa
\Gamma_1(v,m_{\ell}^2,M^2) &=& \frac{1}{2} \Big[ -\,\log R\,+\,
(4-3Y){\cal F}(v,m_{\ell}^2,M^2) \Big]
\nonumber\\
\Gamma_2(v,m_{\ell}^2,M^2) &=& \frac{1}{2} 
\Big(1-\frac{m_{\ell}^2}{v}\Big) 
\Big[ - {\cal F}(v,m_{\ell}^2,M^2)(1-R)
\nonumber\\
&+& \log R \Big]  -
\frac{1}{2}(3-Y){\cal F}(v,m_{\ell}^2,M^2)
\,,
\eeqa
and
\beq
{\cal F}(v,m_{\ell}^2,M^2)\,=\,\frac{2}{\sqrt{R}}\,\frac{X}{1-X^2}\,\log X
\,.
\eeq

\section{Mesonic Loop Functions}
\label{appB}
\renewcommand{\theequation}{\Alph{section}.\arabic{equation}}
\setcounter{equation}{0}

The loop function $H_{PQ} (t)$ \cite{gl852,GL85} is given by 
\begin{equation} 
H_{PQ} (t) =\displaystyle\frac{1}{F_0^2} \bigg[ h_{PQ}^r (t,\mu) 
+ \frac{2}{3}t L_9^r(\mu) \bigg] , 
\end{equation} 
where
\begin{eqnarray} 
h_{PQ}^r (t,\mu) &=& 
\frac{1}{12 t} \lambda 
(t,M_P^2,M_Q^2) \, 
\bar{J}_{PQ} (t) \nn
&&{}+ \frac{1}{18 (4 \pi)^2} (t - 3 \Sigma_{PQ}) \nn
&&{}- \frac{1}{12} \bigg\{ \frac{2 \Sigma_{PQ} - t}{\Delta_{PQ}} [A_P(\mu) 
- A_Q(\mu)] \nn
&& \qquad {}- 2 [A_P(\mu) + A_Q(\mu)] \bigg\}  , 
\end{eqnarray} 
with
\beq
\lambda (x,y,z)  =   x^2 + y^2 + z^2  - 2 ( x y + x z + y z )  , 
\eeq
\beq 
\Sigma_{PQ}  =  M_P^2 + M_Q^2 , \qquad \Delta_{PQ}  =  M_P^2 -
M_Q^2   , 
\eeq
\beq
A_P(\mu)   =   - \frac{M_P^2}{(4 \pi)^2} 
\log \frac{M_P^2}{\mu^2}   , 
\eeq
and
\beqa
\lefteqn{\bar{J}_{PQ} (t)  = 
\frac{1}{32 \pi^2} \Bigg[ 2 + 
\frac{\Delta_{PQ}}{t} 
\log \frac{M_Q^2}{M_P^2} - \frac{\Sigma_{PQ}}{\Delta_{PQ}} 
\log \frac{M_Q^2}{M_P^2} }  \nn
&&{}  - \frac{\lambda^{1/2} (t,M_P^2,M_Q^2)}{t} \nn
&& \times
\log  \left( \frac{[t + \lambda^{1/2} (t,M_P^2,M_Q^2)]^2 - 
\Delta_{PQ}^2}{[t - 
\lambda^{1/2} (t,M_P^2,M_Q^2)]^2 - \Delta_{PQ}^2} \right) \Bigg]   .
\eeqa
The quantity $H_{PQ}(0)$ appearing in the evaluation of $f_+(0)$ is given 
by \cite{gl852}
\beqa \label{null}
H_{PQ}(0) &=& - \frac{1}{128 \pi^2 F_0^2} (M_P^2 + M_Q^2) \,
h_0 \! \left( \frac{M_P^2}{M_Q^2} \right), \nn
h_0(x) &=& 1 + \frac{2x}{1-x^2} \log x.
\eeqa
For the theoretical determination of the slope parameter one needs the 
derivative of the function $H_{PQ}(t)$ at $t= 0$ given by \cite{gl852}
\beqa \label{derH}
\left. \frac{d H_{PQ}(t)}{d t} \right|_{t = 0} &=&
\frac{2}{3 F_0^2} \left\{ L_9^r(\mu) - \frac{1}{128 \pi^2} \log \frac{M_P 
M_Q}{\mu^2} \right\} \nn
&&{} - \frac{1}{192 \pi^2 F_0^2} \, h_1 \! \left( 
\frac{M_P^2}{M_Q^2} \right) , \nn
h_1(x) &=& \frac{x^3 - 3 x^2 - 3 x + 1}{2 (x-1)^3} \log x \nn
&&{}
+\frac{1}{2} \left( \frac{x+1}{x-1} \right)^2 - \frac{1}{3}.
\eeqa

\section{The function $\mbox{\boldmath $I_0(y,z;M_\gamma)$}$ for 
$\mbox{\boldmath $K^0_{\ell 3}$}$}
\label{appC}
\renewcommand{\theequation}{\Alph{section}.\arabic{equation}}
\setcounter{equation}{0}

The analytic result for the integral $I_0(y,z;M_\gamma)$ defined in 
(\ref{I0}) is given by\footnote{Note that the formula for $I_0$ given 
in \cite{gin68} is incorrect even if the Errata are taken into account}

\beqa
\lefteqn{
I_0(y,z;M_\gamma) = \frac{1}{2 \beta} \log 
\frac{1+\beta}{1-\beta}
\log \frac{2 \beta p_{\ell} \! \cdot \! p_\pi}{M_\gamma^2} - \log
\frac{m_{\ell} M_\pi}{M_\gamma^2} 
} \nn
&&{} +\frac{1}{2 \beta} \log \frac{1+\beta}{1-\beta} \log
\frac{2 \beta \gamma (p_{\ell} \! \cdot \! p_\pi)^2 
\left(1-\tau(0)^2\right)^2}
{P \! \cdot \! p_\ell P \! \cdot \! p_\pi} \nn
&&{} +\frac{1}{2\beta} \big[-{\rm Li}_2(\eta_1)+{\rm Li}_2(1/\eta_1)-{\rm 
Li}_2(\eta_2)+{\rm Li}_2(1/\eta_2) \big] \nn
&&{} + \frac{2}{\beta} \bigg[ \log \tau(x_{\rm max}) \log \frac{1-\tau(0) 
\tau(x_{\rm max})}{1-\tau(x_{\rm max})/\tau(0)}  \nn
&&{}  + {\rm Li}_2 \big( \tau(x_{\rm max}) \tau(0) \big)
- {\rm Li}_2 \big( \tau(x_{\rm max}) / \tau(0) \big)  \nn
&&{}  -{\rm Li}_2 \left(\tau(0)^2\right) + \pi^2/6 \bigg] \nn
&&{} + \left( {\rm arcosh} \frac{p_\ell \! \cdot \! p_\pi + x_{\rm 
max}/2}{m_\ell 
M_\pi} \right)^2
- \left( {\rm arcosh} \frac{p_\ell \! \cdot \! p_\pi }{m_\ell M_\pi} 
\right)^2 
\nn
&&{} + \log \frac{4 \, P \! \cdot \! p_\ell \, P \! \cdot \! p_\pi}{x_{\rm 
max}^2} , 
\eeqa
where 
\beqa
\beta &=& \frac{\sqrt{(p_\ell \! \cdot \! p_\pi)^2 - m_\ell^2 M_\pi^2}}
{p_\ell \! \cdot \! p_\pi}, \\
\gamma &=& \frac{P \! \cdot \! p_\ell \, P \! \cdot \! p_\pi}{p_\ell 
\! \cdot \! 
p_\pi} \nn
&\times& \frac{(p_\ell \! \cdot \! p_\pi)^2  -  m_\ell^2 M_\pi^2}{2 \, 
p_\ell \!
\cdot \! p_\pi \, P \! \cdot \! p_\ell \, P \! \cdot \! p_\pi - m_\ell^2 
(P \! \cdot \! 
p_\pi)^2 - M_\pi ^2 
(P \! \cdot \! p_\ell)^2}, \nn
&& \\
P &=& p_K - p_\pi - p_\ell \\
\eta_{1,2} &=& \frac{1 - 2 \gamma \pm \sqrt{\beta^2 + 4 \gamma^2 - 4 
\gamma}}{1 + \beta}, \\
\tau (x) &=& \frac{p_\ell \! \cdot \! p_\pi + x/2 - \sqrt{(p_\ell \! \cdot 
\! 
p_\pi + 
x/2)^2 -m_\ell^2 M_\pi^2}}{m_\ell M_\pi} .
\eeqa

\end{document}